# Habitat connectivity in agricultural landscapes improving multi-functionality of constructed wetlands as nature-based solutions

Clémentine Préau[1*], Julien Tournebize[2], Maxime Lenormand[1], Samuel Alleaume[1], Véronique Gouy Boussada[3], Sandra Luque[1]

[1]INRAE, National Research Institute on Agriculture, Food & the Environment, TETIS Unit, Montpellier, France

[2]INRAE, National Research Institute on Agriculture, Food & the Environment, UR 1462 HYCAR, University Paris Saclay, Antony, France

[3] INRAE, National Research Institute on Agriculture, Food & the Environment, UR Riverly, Villeurbanne

*Correspondence: preau.clementine@gmail.com

**Graphical abstract**

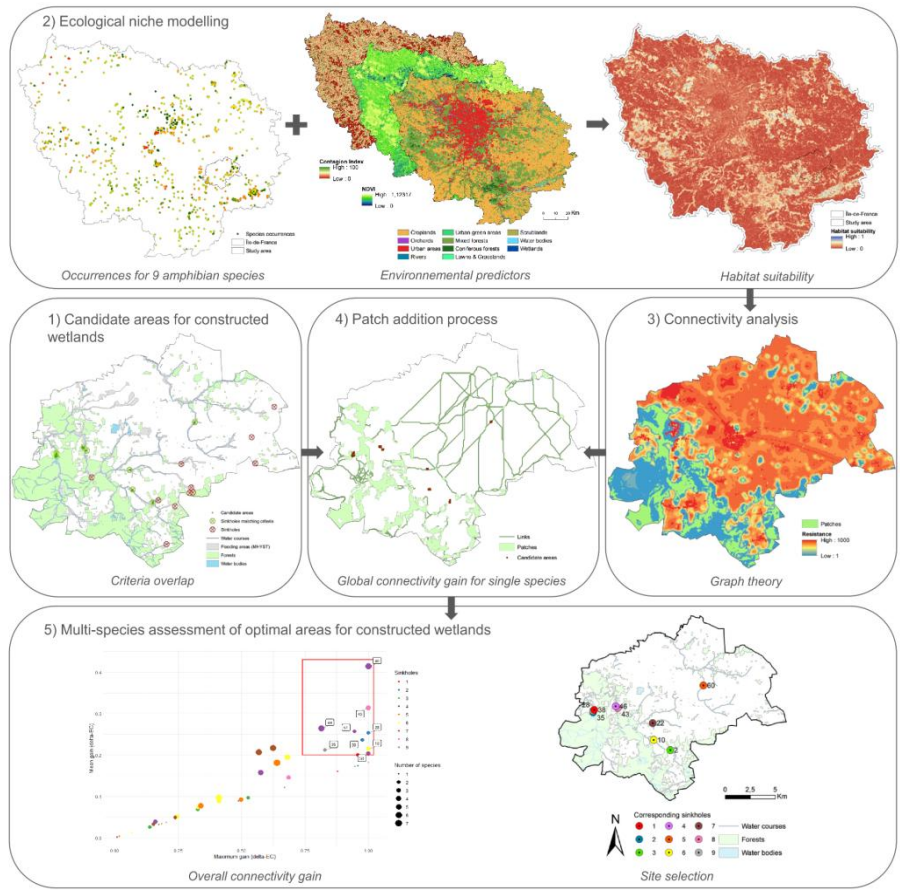

**Abstract**

The prevention of biodiversity loss in agricultural landscapes to protect ecosystem stability and functions is of major importance in itself and for the maintenance of associated ecosystem services. Intense agriculture leads to a loss in species richness and homogenization of species pools as well as the fragmentation of natural habitats and groundwater pollution. Constructed wetlands stand as nature-based solutions (NBS) to buffer the degradation of water quality by intercepting the transfer of particles, nutrients and pesticides between crops and surface waters. In karstic watersheds where sinkholes short-cut surface water directly to groundwater increasing water resource vulnerability, constructed wetlands are recommended to mitigate agricultural pollutants. Constructed wetlands also have the potential to improve landscape connectivity by providing refuge and breeding sites for wildlife, especially for amphibians. We propose here a methodology to identify optimal locations for water pollution mitigation using constructed wetlands from the perspective of habitat connectivity. We use ecological niche modelling at the regional scale to model the potential of habitat suitability for nine amphibian species, and to infer how the landscape impedes species movements. We combine those results to graph theory to identify connectivity priorities at the operational scale of an agricultural catchment area. Our framework allowed us to identify optimal areas from the point of view of the species, to analyze the effect of multifunctional constructed wetlands aiming to both reduce water pollution and to improve amphibian species habitat overall connectivity. More generally, we show the potential of habitat connectivity assessment to improve multifunctionality of NBS for pollution mitigation.

# 1. Introduction

Agriculture intensification has globally been related to biodiversity loss, impacting European landscapes (Burel et al., 1998; Donald et al., 2001; Emmerson et al., 2016; Tsiafouli et al., 2015), in particular through habitat loss, habitat fragmentation and soil and water pollution. Indeed, natural land conversion towards agriculture locally induces loss of habitat connectivity and reduction of flow between populations (*e.g.*, Arntzen et al., 2017), ultimately leading to isolation, decreasing genetic diversity and local extinctions (Cardinale et al., 2012; Frankham, 2005; McGill et al., 2015; Vellend et al., 2013). Intensive agricultural practices are at the center of functional homogenization (Newbold et al., 2015), which is considered to be one of the most conspicuous forms of biotic impoverishment induced by current global changes. Pollution induced by agricultural intensification is also affecting watersheds that are exposed to the transfer of contaminants such as pesticides and fertilizers, leading to the degradation of natural habitats and of water quality (Macary et al., 2014).

Nature-based solutions (NBS) constitute one way to mitigate habitat and water quality degradation in natural systems. Moreover, NBS is an umbrella concept gathering ecosystem-based approaches that provide both human and biodiversity benefits while dealing with specific or multiple societal challenges (Cohen-Shacham et al., 2016, 2019). In agricultural landscapes, semi-natural or artificial buffer zones between crop areas and waterbodies can be used as NBS to mitigate the impacts of agricultural inputs. Landscape elements, such as integrated buffer zones (Zak et al., 2019), vegetative strips (Prosser et al., 2020), riparian buffer zones (Stutter et al., 2019, 2012), vegetated hedges (Lazzaro et al., 2008), or constructed wetlands (Haddis et al., 2020; Metcalfe et al., 2018; Tournebize et al., 2017), can buffer the degradation of water quality by intercepting the transfer of particles, nutrients and pesticides between crops and surface waters. In addition to regulating pollution, buffer areas can improve biodiversity in agricultural landscapes. Vegetated strips and hedges have been shown to improve the abundance and richness of birds and invertebrates, providing habitat and refuge for some species, as well as nesting and foraging sites (see Haddaway et al., 2016). Artificial wetlands can stand for habitat and breeding sites for amphibians (Rannap et al., 2020). For instance, Zak et al (2019) inventoried biodiversity in integrated buffer zones in Sweden and observed colonization by amphibians, small mammals, and invertebrate species. Semi-natural elements can also enhance ecosystem services such as pollination, biological control and soil conservation (Holland et al., 2017). Due to land ownership pressure in intensive agricultural areas, landscape management is difficult to achieve, hence the need for multifunctionality is an environmental issue. The implementation of landscape structures such as constructed wetlands that both mitigate water pollution from agricultural inputs and promote biodiversity meets the objective of multifunctionality. The addition of constructed wetlands enhances hydrological connectivity, even in case of geographically isolated wetlands (McLaughlin et al., 2014), and contributes to the green and blue infrastructure providing corridors or stepping-stones patches enhancing connectivity within the agricultural matrix (Donald and Evans, 2006; EC, 2013). However, such opportunity for biodiversity gain is rarely foreseen in the process of implementation, while early assessment of habitat connectivity would have the potential to increase the multifunctionality of buffer zones, (Dondina et al., 2018; Gippoliti and Battisti, 2017; Hefting et al., 2013) while providing ecosystem services (Mander et al., 2018).

As many studies focus on habitat connectivity by identifying core areas and their links through the landscape matrix (Correa Ayram et al., 2016), landscape graphs have become a popular tool for this purpose (Bergès et al., 2020; Foltête et al., 2020). Among various applications, landscape graphs are widely used to assess the effects of the loss or addition of landscape elements on territorial connectivity and thus to support planning decisions (Foltête et al., 2014). For instance, such methods are used to assess the impacts of urban projects (*e.g.*, Tarabon et al., 2019) or to identify best areas for habitat restoration (*e.g.*, Blazquez-Cabrera et al., 2019; Clauzel and Godet, 2020). Despite growing examples of studies guiding compensation and mitigation of habitat fragmentation and urban development (Bergès et al., 2020), such assessments remain rare in agricultural contexts, whereas promising (Foltête, 2018; Jeliazkov et al., 2014). In particular, this may represent an opportunity to aim for the "no net loss" and "net gain" of biodiversity objectives according to the mitigation practices hierarchy in agricultural landscapes (Calvet et al., 2019). One way of constructing landscape graphs is the use of core habitats and resistance matrix - based upon the facilitating or impeding impact on species movement - identified by ecological niche models (ENMs) (Duflot et al., 2018; Préau et al., 2020). In most cases, ENMs allow correlating environmental variables with species presence and are therefore a good way of obtaining a representation of landscape permeability and, to identify most suitable areas for a specific species (Ziółkowska et al., 2014). ENMs need thorough construction according to the species' ecology to be used in landscape graphs (Godet and Clauzel, 2021). In addition, several studies have shown the interest of accounting for multiple species when evaluating connectivity to cover the needs of taxa depicting various dispersal abilities, and optimizing the efficiency of corridors (*e.g.*, Lechner et al., 2017; Meurant et al., 2018; Sahraoui et al., 2017).

This study focuses on a specific hydrological context of karstic zones characterized by sinkholes intercepting surface water directly to groundwater. Moreover, due to intensive subsurface drained farming, agricultural pollutants are rapidly transferred making the groundwater highly vulnerable. We propose a methodology to identify optimal areas for pollution mitigation based on the implantation of buffer zones, namely constructed wetlands (CW, also called artificial wetlands) from the perspective of habitat connectivity. Using ENMs and graph theory, our assessment is based on a multispecies group of amphibians. We use an overview of the potential habitat suitability at a regional scale as a basis to model local scale connectivity of the study site and to identify candidate CW implementation areas providing the best optimization opportunity to add multifunctionality through biodiversity conservation in addition to pollution mitigation. Our approach, within the framework of NBS, explicitly accounts for habitat connectivity in the decision process of CW implementation at the operational scale of an agricultural catchment area.

## 2. Material and methods

### 2.1 Study site and candidate areas for constructed wetlands

The study site is in the administrative region Ile-de-France. It is concentrated on an hydrological catchment called Ru d'Ancoeur in Brie, covering an area of 224 km² and dominated by intensive agriculture (Fig. 1). The study area is complemented by a tile drainage system, mainly built during the 60s' and 80's, that helped control water excess in winter in order to fulfill yield objectives on hydromorphic agricultural plots. For this reason, surface run-off within the study site is not significant compared to undrained soil, and the most relevant method of mitigating pollution therefore seems to be the implementation of artificial buffers zones such as constructed wetlands (CW) to capture drainage water (Tournebize et al., 2017). Depending on seasonality, environmental parameters and wetland design, CW can have a high denitrification and pesticide retention and degradation potential (Haddis et al., 2020; Tournebize et al., 2017, 2015). The study site also includes sinkholes in karst areas, which are preferential infiltration areas towards the groundwater, identified and validated by the association Aqui'Brie in charge of water quality management on the site (https://www.aquibrie.fr/). Consequently, CWs can be implemented close to watercourses and upstream of sinkholes to capture pollution flows before infiltration into the groundwater which is a crucial resource for drinkable water in Paris suburb (1.5 M of inhabitants).

The workflow of analysis is provided through graphical representations of steps detailed in Appendix A. Potential areas for the implementation of CW were targeted through a set of criteria. Because subsurface pipes network maps are not available, we propose 1) to mitigate surface water upstream of sinkholes location 2) to consider the flooded areas as proxy of hydrological nodes suitable for restoration of wetlands. As a result, we identified sinkholes that were both close to a watercourse (<150m) and located in a flooding area. The maximum extent of flooding areas in the site had previously been modeled through a flood inundation model called MHYST (see Letournel et al., 2021; Rebolho et al., 2018). This resulted in the identification of nine sinkholes (Fig. 1). Then, we identified areas that were both upstream, at a maximum distance of 150 m from one of these nine sinkholes, and outside already existing areas of forest or waterbodies. This resulted in 60 pixels of 50x50 m as candidate areas for CW (Fig. 1).

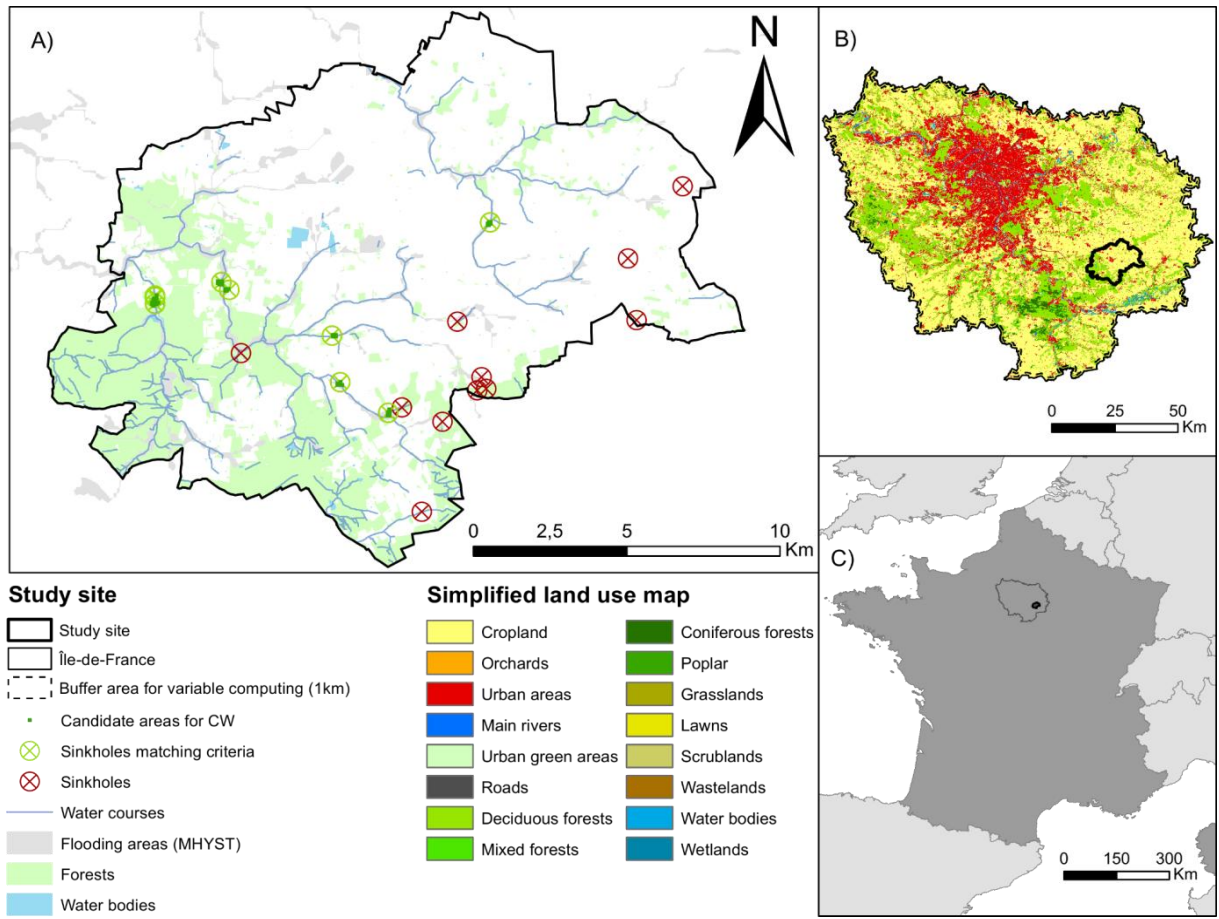

**Fig. 1.** Localization of candidate areas for the implementation of CW within the study site (A). Localization of the study site within the region Ile-de-France, used to train ENMs (land use map) (B) and in France (C).

**2.2 Species and environmental data**

We obtained species data from the regional Île-de-France naturalist database CETTIA [dataset](CETTIA) and the National Inventory of Natural Heritage [dataset](INPN). We selected a set of nine amphibian species for the period from 2010 to 2015: *Alytes obstetricans*, *Bufo bufo*, *Epidalea calamita*, *Hyla arborea*, and *Rana temporaria* from the anurans order, as well as *Ichthyosaura alpestris, Lissotriton vulgaris, Salamandra*

*salamandra* and *Triturus cristatus* from the urodeles order (see details on species data in Appendix B). We spatially rarefied presence points of the species within a distance of 50 m around each point to minimize sampling bias and point clustering (Boria et al., 2014). We compiled several land cover /land use and remote-sensing databases on the Île-de-France region containing the study area: ECOMOS and MOS [dataset](Institut Paris Région a, b), National Institute of Geographic and Forest Information [dataset](IGN), Theia-land [dataset](CESBIO). We computed 206 environmental variables considering 1 km of buffer area, at the resolution of 50x50 m (Appendix C).

## 2.3 Ecological niche modelling

We used the Biomod2 platform for ensemble modelling (Thuiller et al., 2009) under R software version 3.6.0. We trained models across the extent of the Île-de-France region to account for a large environmental gradient (Fig. 1B). We applied a selection procedure proposed by Leroy et al. (2014) to select a set of uncorrelated variables based on Pearson correlation ($r<0.7$) and variable importance for each species. The procedure was repeated three times in a row and the ten more important variables were selected for each species (Brun et al., 2020) (see Appendix B for variable description and retained variables per species). We ran the models with nine algorithms, grouping generalized additive model (GAM), generalized linear model (GLM), and multivariate adaptive regression splines (MARS), artificial neural networks (ANN), flexible discriminant analysis (FDA), classification tree analysis (CTA), generalized boosting models (GBM), random forest (RF), and MAXENT. We ran the models with five sets of pseudo-absences (PA) selected at a minimal distance of 50 m and at a maximal distance of 2 km from our occurrences (disk method), as this method has shown to perform well with few presence data (Barbet- Massin et al., 2012). We generated the same number of PA as for presence data for each species. We used three runs per model and equal weight for presence and PA. This resulted in a total of 450 models for each species. We made 3-fold cross-validation of our models by randomly splitting the observation dataset into 70% for training and 30% for the evaluation by the area under the ROC curve (AUC-ROC, Hanley and McNeil (1982), True skill statistic (TSS, Allouche et al. (2006), and Boyce Index (Boyce et al., 2002; Hirzel et al., 2006). We used ensemble modelling to display central tendency across modelling algorithms and included all runs (Araújo and New, 2007).

## 2.4 Connectivity analysis and patch addition process

We used graph theory through Graphab v2.6.1 (Foltête et al., 2012) to evaluate the potential of each candidate area for constructed wetlands (CW) based on connectivity analysis for amphibian species. At this step, we assume that CW, in drained areas, are suitable habitats for amphibians. We computed the subsequent analysis separately for each amphibian species, for the spatial extent of the study area (Fig.1A). Using the output Habitat suitability index (HSI) maps of ENMs, we defined the nodes of the graph as suitable patches after binary discrimination of suitable against unsuitable areas using the $10^{th}$ percentile threshold (Capinha et al., 2013; Pearson et al., 2007). We built a complete graph (*i.e.*, all links between all patches area accounted for) considering 8 neighbors, ignoring links crossing patches. We defined landscape resistance as cumulative cost, from the continuous HSI maps such as: resistance =1 when HSI was ≥ $10^{th}$ percentile threshold; and resistance = e(ln(0.001)/$10^{th}$ percentile threshold x HSI) ×$10^3$ when HSI was < $10^{th}$ percentile threshold (Duflot et al., 2018). We set patch capacity as the mean of habitat suitability within patch areas. We converted metric distance to cost-unit distance using the maximal dispersal distance of species (*i.e.*, 1000 m for urodeles and 5000 m for anurans, see Appendix B for references and resulting cost-unit distances). We calculated a global metric: the Equivalent Connectivity (EC) index for each graph (Clauzel and Godet, 2020; Saura et al., 2011). This metric ranges from 0 to the sum of patch capacities. We assessed the relevance of each of the 60 candidate pixels for CW implementation, regarding each species, using the patch addition process initially developed by Foltête et al. (2014) and, for instance, recently used in Clauzel and Godet (2020) and Tarabon et al. (2021). The capacity of CW candidates was set as the value of HSI at its location. By iteratively calculating the value of the EC index and the associated gain after the addition of a CW candidate that would constitute a new patch (delta-EC), this method allowed to optimize the selection of patches that provided the greatest gain. Candidate pixels already belonging to a patch were excluded from the analysis. This allowed to rank new patches (*i.e.*, candidate areas for CW) validated at each step of the process according to their potential for maximizing connectivity (*i.e.*, the EC metric) for each species.

**2.5 Optimal areas for CW regarding overall connectivity**

We standardized the gains (delta-EC) associated with each candidate area by the maximum gain obtained for each species. For each candidate area, we then estimated the overall connectivity potential by calculating the mean and maximum values of gain over the nine species. We ranked the candidate areas for CW implementation according to the overall connectivity gain considering two scales. The scale of the study area (*i.e.*, site scale) included all candidate areas, while the sinkhole scale encompassed the candidate areas associated with each sinkhole. At the site scale, we identified the nine top candidate areas with the highest mean and maximum values as areas most suitable for CW implementation.

We then calculated Jaccard's distances between the candidate areas to assess the similarity of species representation. We also identified the number of species represented by each of the nine top candidate areas. Based on these results we were able to identify the top 1 candidate area for CW implementation. At the sinkholes scale, we identified the top candidate area per sinkhole from the mean and maximum values. We identified the number of species represented by the top candidate area of each sinkhole. We computed the difference of mean and maximum gain between the identified top 1 candidate area and the other top candidate areas identified at the two scales, and the additional number of species showing connectivity gain, in order to guide the identification of the subsequent candidate areas suitable to later CW implementation.

## 3. Results

### 3.1 Ecological niche modelling and connectivity analysis

Ecological niche models globally provided good scores of TSS (from 0.619 for *I. alpestris* to 0.961 for *E. calamita*); AUC (from 0.876 for *I. alpestris* to 0.994 for *E. calamita*) and Boyce Index (from 0.718 for *E. calamita* to 0.987 for *L. vulgaris*). Details for evaluation metrics, binary thresholds and variable importance in ensemble models are available in Appendix B. The most important variable identified for each species is available in Table 1. The surface area identified as suitable according to the binary thresholds, and as suitable patches for connectivity analysis, varied from 0.05 % of the study area for *E. calamita*, which was mainly associated to wasteland areas, to about 27 % of the study area for *I. alpestris*, which was closely associated to mixed and deciduous forests. Results for *T. cristatus* and *B. bufo* also showed > 20 % of the study area as suitable, whereas that percentage was between 9.6 % and 13.7 % for the other species.

Regarding connectivity analysis, we found the study area to be inequality connected according to the species. Indeed, results for global EC metric were the lowest for *E. calamita* with 1.906, while it reached the highest values for *T. cristatus*, *H. arborea* and *B. bufo* (respectively 109.959, 171.159 and 195.308). Maximum dispersal distances used in connectivity analysis are listed in Table 1. The potential of candidate areas to increase connectivity for the nine species was also uneven, resulting in a different number of potential new patches for each species. It ranged from 9 validated new patches for *H. arborea*, *I. alpestris* and *T. cristatus* to 19 for *E. calamita*. Detailed results on graph analysis are available in Appendix B.

**Table 1.** Species most important variable resulting from ENMs, and dispersal distance used in analysis.

| *Species* | *Most important variable identified in ENMs* | Maximum dispersal distance used for analysis in meters |
|---|---|---|
| *Alytes obstetricans* | Color index | 5000 |
| *Bufo bufo* | Distance to water bodies | 5000 |
| *Epidalea calamita* | Distance to wastelands | 5000 |
| *Hyla arborea* | Distance to water bodies | 5000 |
| *Ichthyosaura alpestris* | Distance to mixed and deciduous forests | 1000 |
| *Lissotriton vulgaris* | Distance to water bodies | 1000 |
| *Rana temporaria* | Surface area of deciduous forests within 200m | 5000 |
| *Salamandra salamandra* | Surface area of deciduous forests within 200m | 1000 |
| *Triturus cristatus* | Distance to water bodies | 1000 |

### 3.2 Optimal areas for CW

We found 5 shared top candidate areas between the two scales. Considering the site scale (*i.e.*, all sinkholes together), the top 9 candidate areas with the highest overall connectivity were: 46; 43; 44; 51; 28; 30; 35; 10 and 50 (Fig. 2A). Considering the sinkholes scale, the top 9 candidate areas were: 38; 28; 2; 46; 60; 10; 22; 43 and 35 (Fig. 3). Top candidate sites for constructed wetlands (CWs) implementations 22 and 44 were the most similar regarding binary estimation of connectivity gain (0 when gain was null and 1 when gain was positive). We found *H. arborea* and *A. obstetricans* to be the most similar regarding binary estimation of connectivity gain (Appendix D).

The candidate area 46 showed connectivity gain for 6 species and the highest values of mean and maximum gain. As a result, we identified the candidate area 46 as the most suitable candidate area for CW implementation. It is related to sinkhole 4, close to the western limit of the forested part of the study area. It is next to a forest patch (Fig. 4). Connectivity gains with candidate area 46 were the most important for *A. obstetricans*, *E. calamita*, *I. alpestris* and *L. vulgaris*. Conversely, connectivity for *T. cristatus* and *H. arborea*, which already showed high global EC values, was not improved with the addition of the candidate area 46 (Fig. 2B).

After identifying the most suitable area for the implementation of CW, Fig. 2C can guide the selection of candidate areas where CW should be implemented afterwards, based on a comparison of the gains against the candidate area 46. Indeed, Fig. 2C shows the difference of mean and maximum connectivity gain of all identified top candidate areas (Fig. 2A and Fig. 3) compared to candidate area 46, and the number of additional species gaining connectivity (*i.e.*, species not already accounted for at candidate area 46). For instance, candidate area 43 showed potential for additional gain, with the closest values of mean and maximum connectivity gain to those of the candidate area 46, but it provided additional gain for only one species not already accounted for at candidate area 46 (Fig. 2C). Moreover, the new species at candidate area 43, *S. salamandra*, showed low connectivity gain at this location (Fig. 2B). In contrast, the candidate area 60 showed gain for three more species but the overall gain was actually the lowest compared to the candidate area 46 and, among additional species, *T. cristatus* and *H. arborea* already showed some of the highest global EC. As a result, other top candidate areas could provide more important additional gain for other species than candidate area 46. Indeed, candidate area 2 showed important additional gain only for *R. temporaria*, and candidate area 38 showed important additional gain for *S. salamandra*. It would also allow more variation regarding the location of CW, with connection to farther sinkholes (Fig. 4).

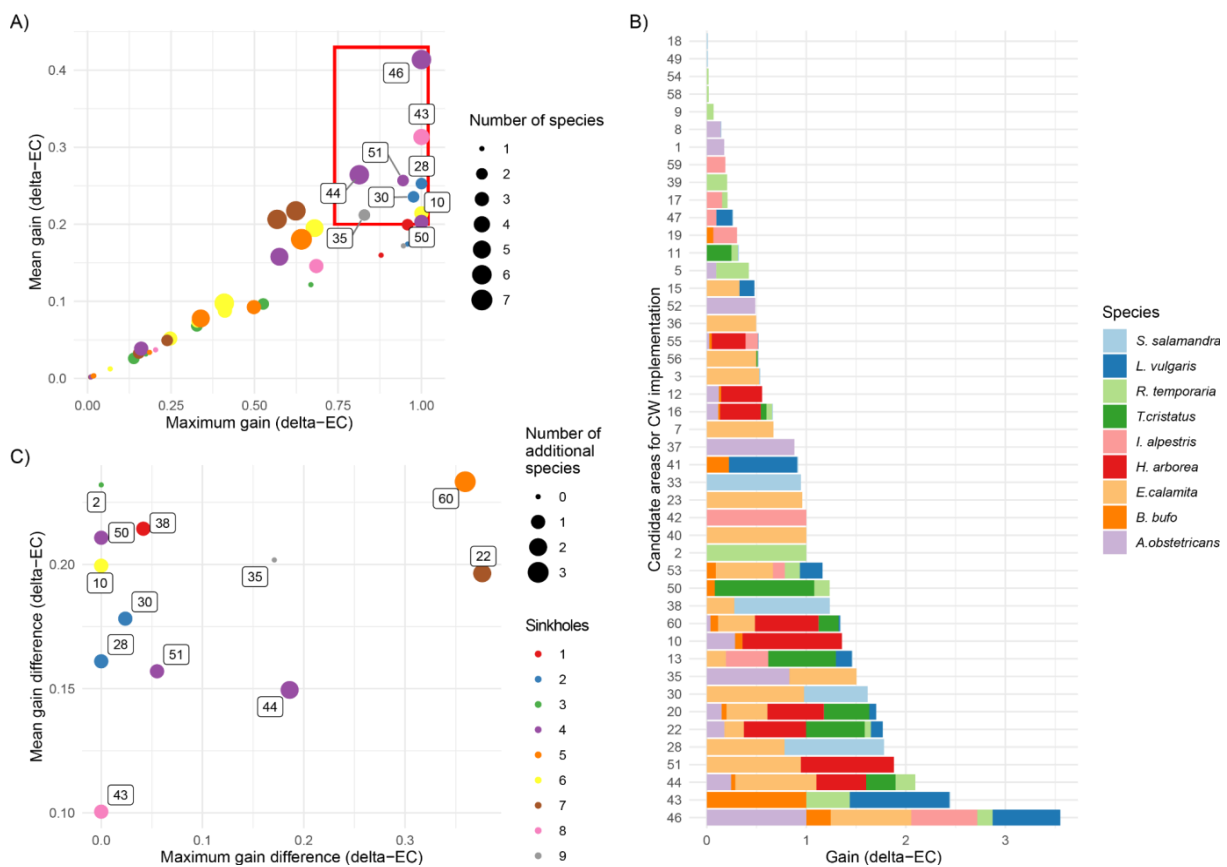

**Fig. 2.** A) Distribution of candidate areas for constructed wetland (CW) at site scale according to the values of maximum and mean gain (delta-EC). The top 9 candidate areas are identified in the red square. The labels indicate the number associated with the candidate area. Colors identify the corresponding sinkhole. The size of the dots indicates the number of species for which the candidate area provides a connectivity gain. B) Connectivity gain (delta-EC) for individual species potentially provided by each candidate area for CW implementation. C) Difference from the identified top candidate area 46 for all other identified top candidate areas according to the values of maximum and mean gain (delta-EC). The labels indicate the number associated with the candidate area. Colors show the corresponding sinkhole. The size of the dots indicates the number of additional species accounted for compared to the candidate area 46.

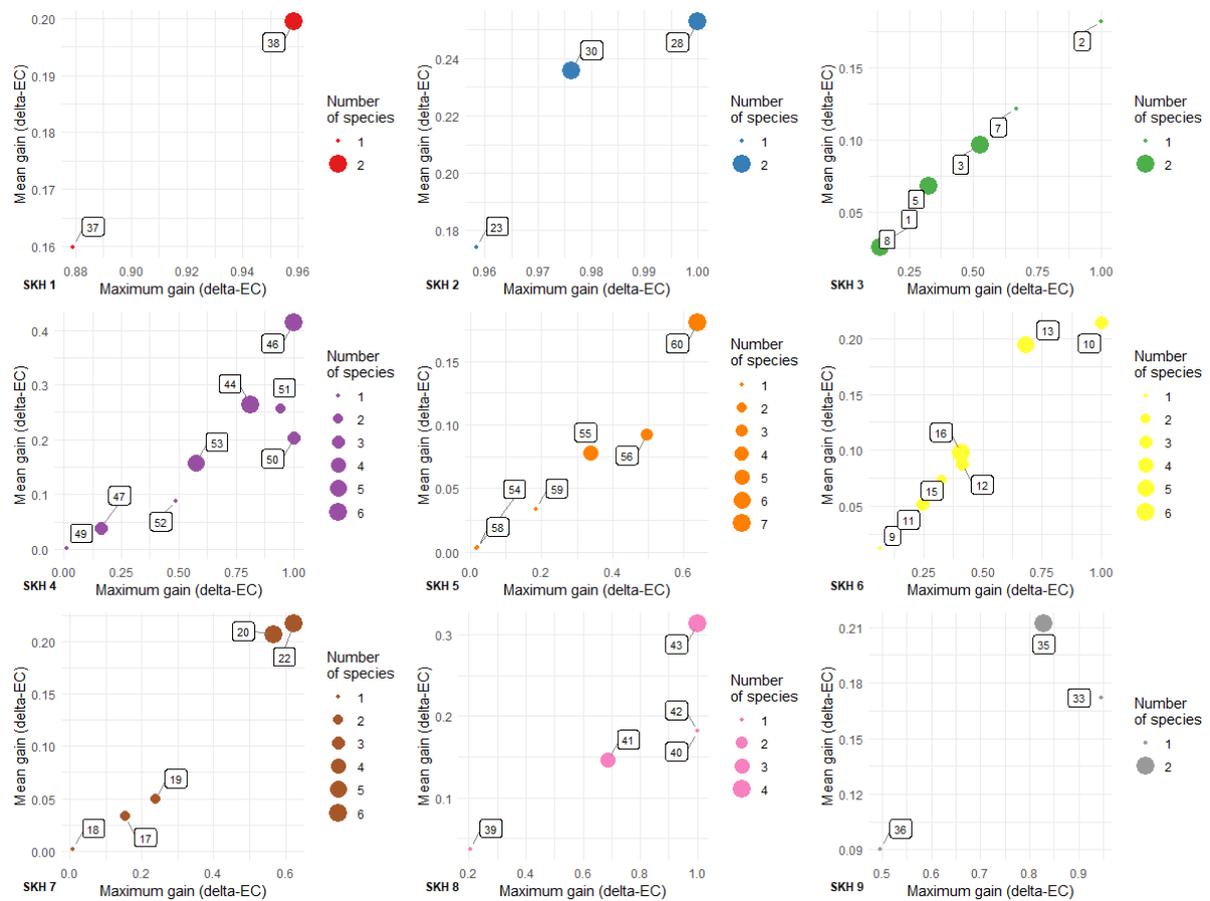

**Fig. 3.** Distribution of candidate areas for constructed wetland (CW) implementation at sinkholes scale, according to the values of maximum and mean gain (delta-EC). The labels indicate the number associated with the candidate CW. The size of the dots indicates the number of species for which the candidate CW provides a connectivity gain. SKH stands for sinkhole.

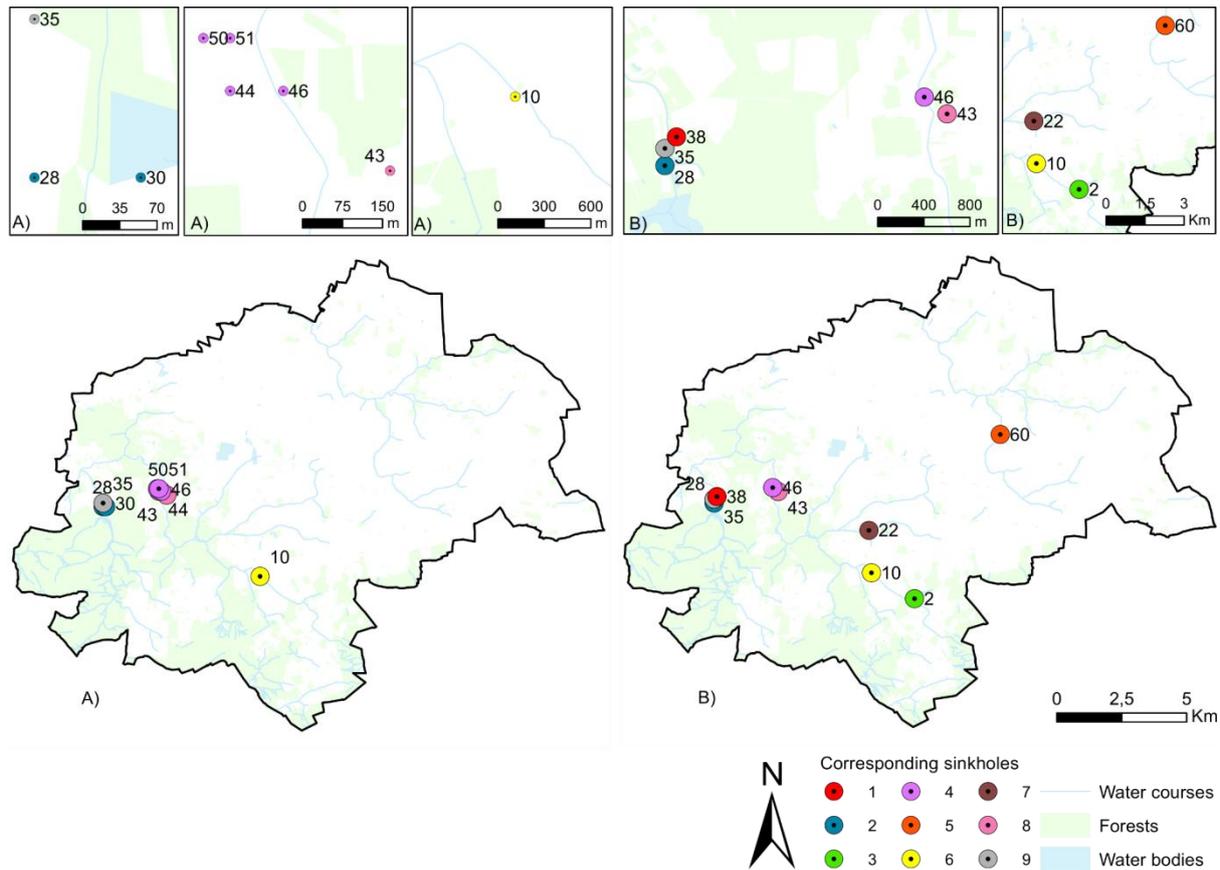

**Fig. 4.** Spatial distribution of the top 9 candidate areas for constructed wetland (CW) when A) considering all candidate areas of the study area (*i.e.*, site scale) or B) when considering the sinkholes separately (*i.e.*, sinkholes scale). The labels indicate the number associated with the candidate CW while the color refer to the corresponding sinkhole.

## 4. Discussion

Our methodology allowed to identify an optimal area to create a multifunctional constructed wetlands (CW) aiming to both reduce water pollution and improve amphibian species overall connectivity within the study area. The protection of water induces an additional constraint for CW implementation by favoring the upstream of sinkholes, as preferential pollutant pathway from surface water to groundwater. The combination of ENMs with graph theory helped identify connectivity priorities for the nine species and to rank candidate areas identified as suitable for reducing water pollution by means of a CW, insuring net biodiversity gain in a territory dominated by intensive agriculture.

While some species showed relatively high overall connectivity values, according to our results, the candidate area 46 appeared to be the most appropriate for the implementation of a CW as it would improve connectivity for additional species. However, this particular area was not convenient to improve connectivity of all species. Clauzel and Godet (2020) studied the possibility of restoring the connectivity of seven amphibians and a reptile at places spaced by 100 m throughout the whole region Ile-de-France. They did not find an area that would allow a compromise for the eight study species combined, which corroborates our results at a larger spatial scale. This is because although all study species can use CW as habitats for reproduction, they have different profiles of suitable habitat and of initial state of connectivity. At the regional scale, Clauzel and Godet (2020) found that areas close to both forest and agricultural areas were the most suitable for improving multi-species connectivity. This supports our choice of candidate area 46 which lies at the boundary between the forested part of our study area and the heavily agricultural part.

Subsequently to area 46, other top candidate areas offer opportunities for CW implementation to meet different objectives. For instance, wetland construction at candidate area 43 could strengthen connectivity improvement especially for *B. bufo* and *L. vulgaris*, which already show high overall connectivity compared to other species and thus could be more able to sustain through the agricultural landscape. In addition, CW implementation could also improve species richness by enhancing connectivity for species missing or showing low connectivity gain at area 46. This would lead to select areas regardless of the smaller number of species represented, such as candidate area 38 to favor connectivity for *S. salamandra* or candidate area 2 to favor *R. temporaria*. The candidate area 60, which is of interest for the connectivity of seven of the nine study species, could also be seen as a means of promoting species richness. However, it provides only small overall connectivity gains. As this area is far from the western forest patches, it would be worthwhile to improve connectivity at first in areas close to refuge habitats, then to gradually reconnect the two parts of the study area. Indeed, conservation efforts are more effective to amphibians when carried out near pre-existing breeding habitats rather than on isolated habitats, as this improves the functioning of metapopulations (Jeliazkov et al., 2014; Peterman et al., 2018). This may require the addition of new patches at locations that have not been prioritized for water quality issues, in order to complete the network, and provide stepping-stones promoting connectivity among remaining populations and recolonization (Semlitsch, 2002).

While CWs have shown efficiency as NBS towards water treatment (Tournebize et al., 2017; Vymazal and Březinová, 2015), several studies have also reported their ability to enhance biodiversity in various agricultural landscapes, for instance for invertebrates (Becerra-Jurado et al., 2012) or birds (Letournel et al., 2021; Strand and Weisner, 2013). Our assessment of connectivity gain is based on the hypothesis that CWs can provide suitable breeding habitat for amphibians (Rannap et al., 2020). However, in different contexts, it has been shown that amphibian occurrence in a suitable habitat is influenced by the composition and configuration of the surrounding landscape (Boissinot et al., 2019; Ferrante et al., 2017; Guerra and Aráoz, 2015; Ruso et al., 2019; Sawatzky et al., 2019). We took habitat suitability into account when assessing the connectivity gains of the candidate areas by defining patch capacity as the HSI index in the 50x50 m grid during the patch addition process. Furthermore, while intensively drained and cultivated agricultural land threaten the survival of amphibians, particularly through exposure to pollutants and desiccation (Brühl et al., 2013; Cosentino et al., 2011), the tolerance for such landscape has shown to be species specific (Baker et al., 2013; Rannap et al., 2020). CWs are commonly used to reduce the impact of agricultural runoff and drainage systems, then being the receptacle of pollutants such as pesticides and nitrogen (Vymazal and Březinová, 2015). Amphibian species colonizing such CWs are therefore exposed to pollution, that can lead to sublethal effect, such as disturbances in larval development and metamorphosis, and susceptibility to diseases or parasites (Rannap et al., 2020; Smalling et al., 2015; Swanson et al., 2018). Considering the risk of CWs to becoming ecological traps for amphibians (Zhang et al., 2020), Rannap et al. (2020) recommend minimizing the concentration of accumulated nutrients in the plant biomass by removing sediment and thick macrophyte cover during late autumn, after larval metamorphosis of all species. Further ecotoxicological studies are needed to assess synchronism and antagonism in seasonal pesticides transfer and life cycle of species such as amphibians. Moreover, while some species' tadpoles can show defense against fish predation, such as *B. bufo* (Hossie et al., 2017; Üveges et al., 2019), the presence of fish remains limiting for the reproduction and population dynamics of most species (Hartel et al., 2007; Rannap et al., 2009; Schmidt et al., 2021) and should be avoided in CWs (Jeliazkov et al., 2014; Rannap et al., 2020). In order to promote the colonization by amphibians, CWs and immediate vegetative buffer zones should provide various microtopographic features and gradual slopes, allowing for calling, oviposition, foraging, thermoregulation, and refuge, as well as low soil compaction and diverse hydroperiods as sites for breeding during wet and dry years (Drayer and Richter, 2016; Semlitsch, 2002).

## 5. Conclusions

Our approach allowed to target a set of priority areas based on a set of criteria for the establishment of artificial wetlands that would be suitable to several amphibian species as well as playing a part in the regulation of pollution from agricultural inputs. In addition to connectivity gain for amphibians, the multifunctionality of CWs within the study area would improve the overall ecological network (Clauzel and Godet, 2020) and thus provide a set of ecosystem services (Keesstra et al., 2018; Liquete et al., 2016; Thorslund et al., 2017), as well as increasing the potential for adaptation to climate change (Chausson et al., 2020). However, the benefits of CWs should not overtake the value of existing natural landscape features, which should be a conservation priority (Magnus and Rannap, 2019). As a result, land sparing (*i.e.*, separation of production fields and natural habitats) and land sharing (*i.e.*, environmentally friendly agricultural systems) measures can be combined, through improved connectivity of the landscape matrix, to enable the conservation of both generalist and specialist species and to facilitate species movement and metapopulation functioning (Grass et al., 2019). The analysis of connectivity based on species distribution, patch addition process, and identification of optimal areas for NBS is easily replicable and transposable to other agricultural contexts. This approach can also be extended to other species depending on the landscape elements to be added, and thus contribute to the challenges of regaining biodiversity within agricultural landscapes.

**Acknowledgements**

This research would not have been possible without the financial support of the French Agency for Biodiversity (AFB), and the French Ministries of Agriculture and Environment to the SPIRIT project within the framework of the French research program "Leviers territoriaux", and Life+ ARTISAN project (LIFE18 IPC/FR/000007).

We thank the structures and the observers associated with the data extracted from the regional tool Cettia îdF, as well as of the SINP (Information system of the Inventory of the Natural heritage) provided by the INPN (National Inventory of the Natural heritage). We thank the association Aqui'Brie.